\def\eV{{\rm eV}}
\def\ue3{\left| U_{e3} \right|}

\def\be{\begin{equation}}
\def\ee{\end{equation}}

\documentstyle[epsfig,12pt]{article}
\begin{document}
\baselineskip=22 pt
\setcounter{page}{1}
\thispagestyle{empty}
\topskip  -2.5  cm
\begin{flushright}
\end{flushright}
\vspace{0.2 cm}

\centerline{\LARGE \bf  Model of Mass Varying Neutrinos  in SUSY}
\vskip 0.8 cm
\centerline{{\large Ryo Takahashi$^{a}$}$\!\!\!$
\renewcommand{\thefootnote}{\fnsymbol{footnote}}
\footnote[1]{e-mail:  takahasi@muse.sc.niigata-u.ac.jp} \
{\large and  Morimitsu Tanimoto$\,^{b}$}$\!\!\!$
\renewcommand{\thefootnote}{\fnsymbol{footnote}}
\footnote[6]{e-mail: tanimoto@muse.sc.niigata-u.ac.jp}}

\vskip 0.5 cm

\centerline{$^a\!$ Graduate School of  Science and Technology,
 Niigata University,  950-2181 Niigata, Japan}

\centerline{$^b\!$ Department of Physics,
Niigata University,  950-2181 Niigata, Japan}

\vskip 2 cm
\centerline{\bf ABSTRACT}\par
\vskip 0.2 cm
We discuss  the mass varying neutrino scenario in the supersymmetric theory.
 In the case of the model with the single superfield, one 
needs the soft SUSY breaking terms or the $\mu$ term. However,
 fine-tunings of some parameters are required to be consistent with 
 the cosmological data. 
In order to avoid the fine-tuning, we discuss the model with two superfields,
 which is consistent with the  cosmological data. 
 However, it is found that the left-handed neutrino  mixes
 with the neutrino of the dark sector maximally.
Adding a right-handed neutrino, which does not couple to the dark sector,
 we obtain a favorable  model  in the phenomenology of the neutrino 
experiments.
 In this model, the deceleration of the cosmological expansion
 converts to the acceleration  near $z\simeq 0.5$.
The speed of sound $c_s$  becomes imaginary   
if we put $w_0=-0.9$, which corresponds to $m_\nu^0=3.17\ \eV$.
On the other hand,   if we take $w_0=-0.998$, which leads to
$m_\nu^0=0.05\ \eV$,  $c_s^2$  becomes positive  since
$w$ evolves rapidly near the present epoch in our model.

\newpage

\section{Introduction}

One of the most challenging questions in both cosmological and 
particle physics is the nature of the dark energy in the Universe.
At the present epoch, the energy density of the Universe is dominated 
by a dark energy component,
whose negative pressure causes the expansion of the Universe to accelerate.
 In order to clarify the origin of the dark energy, one has tried to 
understand the connection of the dark energy with particle physics.

Recently, Fardon, Nelson and  Weiner  \cite{Weiner}  proposed an idea
 of the mass varying neutrinos (MaVaNs), 
in which the neutrino couples to  the dark energy.
The variable  neutrino mass  was considered at first in 
\cite{Yanagida}, and was discussed for neutrino clouds \cite{McKellar}.
  However, the renewed  MaVaNs scenario \cite{Weiner}  has tried to make a
 connection between neutrinos and  the dark energy. In this  scenario,
an unknown scalar field which is called ``acceleron'' is  introduced,
and then, the neutrino mass becomes a dynamical field.

The acceleron field sits at the instantaneous minimum of its potential, 
and the cosmic expansion only modulates this minimum through changes 
in the  neutrino density.
Therefore, the neutrino mass is given by the acceleron field
and changes with the evolution of the Universe.
The cosmological parameter $w$ and
 the dark energy also evolve with  the neutrino mass.
Those evolutions depend on a model of the scalar potential strongly. Typical 
examples of the potential have been discussed by Peccei \cite{Peccei}. 

The  MaVaNs scenario leads to interesting  phenomenological results.
 The neutrino oscillations may be a probe of the dark energy 
\cite{Kaplan,Mini}.
 The baryogenesis \cite{Wang,Bi,Gu}, the cosmo MSW effect of  neutrinos 
\cite{Hung} and the solar neutrino \cite{SUN1,SUN2} 
have been studied in the context of this  scenario.
 Cosmological  discussions of  the  scenario are also presented 
\cite{Zhang,Right,stability,WZ,Supernova}.

In this paper,
 we study the MaVaNs scenario in the supersymmetric theory and 
construct   models which are consistent with the  current cosmological data 
\cite{Data}.
 We discuss the dark energy in the some cases of the superpotential.
Then,  we present the numerical results for the  evolution of 
the neutrino mass and  $w$.
In section 2 and 3, we study the superpotential with the single superfield
 and the double  superfields, respectively.
The section 4 devotes to the discussions and summary.

\section{The single superfield model}

The simplest assumption of the MaVaNs with the supersymmetry 
is to introduce a single chiral superfield $A$, which is a singlet under 
the gauge group of the standard model.   
The  superfield $A$  couples to  the left-handed lepton  superfield $L$.
 In this  framework, we discuss  three cases of the superpotential.

\subsection{The simplest model $W=\frac{\lambda}{3}A^3+m_DLA$}

We suppose  the dark sector with the superpotential
\begin{equation}
W=\frac{\lambda}{3}A^3+m_DLA \ ,
\label{spotential1}
\end{equation}
\noindent where $\lambda$ and $m_D$ are a coupling constant and 
a mass parameter, respectively.  
The scalar and spinor components of $A$  are $(\phi_a,\psi_a)$,
and  the scalar component $\phi_a$ is assumed to be the acceleron.
The second term of the right hand side $m_D LA$ in eq.(\ref{spotential1})  
is derived from the Yukawa coupling  $y LAH$ with $y\langle H\rangle=m_D$,
where $H$ is  the Higgs doublet.  Assuming the vanishing vacuum 
expectation value of the left-handed slepton,  the superpotential
 of eq.(\ref{spotential1}) is enough to discuss the dark energy,
because  only the scalar potential 
of the acceleron and the neutrino energy density contribute to 
the dark energy in the MaVaNs  scenario.
We  omit other terms, 
which does not couple to $A$,  in our   superpotential.

Then, the scalar potential for $\phi _a$ is given by
\begin{equation}
V(\phi _a)=\lambda ^2|\phi _a|^4+m_D^2|\phi _a|^2 \ .
\label{potential1}
\end{equation}
We can write down a Lagrangian density from eq.(\ref{spotential1}),
\begin{equation}
{\mathcal L}=m_D\nu _L\psi _a+2\lambda\phi _a\psi _a\psi _a \ .
\end{equation}
 This means that the dark sector interacts with the standard electroweak
sector only through neutrinos. By solving the eigenvalue equation of 
the $2\times 2$ mass matrix,
 \begin{equation}
    \left( \begin{array}{ccc}
     0    & m_D        \\
     m_D &  2\lambda \phi_a    
    \end{array} \right) \ ,
 \end{equation}
\noindent
  $\phi_a$ is given in terms of the neutrino mass,
\begin{equation}
 \lambda \phi_a=\frac{m_\nu}{2}-\frac{m_D^2}{2 m_\nu}\ .
\label{mass}
\end{equation}
\noindent
Using this relation, the scalar potential of eq.(\ref{potential1}) is
given in terms of the neutrino mass $m_\nu$ as follows:
\begin{equation}
V(m_\nu )=
\frac{1}{16\lambda ^2}\left ( m_\nu-\frac{m_D^2}{m_\nu}\right )^4
+\frac{m_D^2}{4\lambda ^2}\left ( m_\nu-\frac{m_D^2}{m_\nu}\right )^2,
\label{potential2}
\end{equation}
where, for simplicity, we take the scalar field real.

In the MaVaNs scenario, there are two constraints on the scalar
potential. The first one comes from the observation of the Universe, 
which  is that the present dark energy density is about $0.7\rho_c$, 
$\rho _c$ being  a critical density.
 Since the dark energy is assumed  to be the sum of the energy densities 
 of the neutrino and the scalar potential
\begin{equation}
 \rho_{\mbox{dark}}=\rho _\nu +V(\phi_a(m_\nu ))  \ ,
\end{equation} 
\noindent the first constraint turns to 
\begin{equation}
\rho _\nu ^0+V(\phi _a^0(m_\nu ^0))=0.7\rho _c \ ,
\end{equation}
where ``0'' represents a value at the present epoch and
$70\%$ is taken for the dark energy in the Universe. 

The second one
comes from the fundamental assumption in this scenario, which  is that
$\rho _{\mbox{dark}}$ is stationary with respect to variations in the
neutrino mass. This assumption is represented by
\begin{equation}
\frac{\partial\rho _\nu}{\partial m_\nu}+
\frac{\partial V(\phi_a(m_\nu ))}{\partial m_\nu}=0 \ .
\label{stationary1}
\end{equation}
For our purpose it suffices to consider the neutrino mass as a
function of the cosmic temperature \cite{Peccei}. Then the stationary
condition eq.(\ref{stationary1}) turns to \cite{Peccei}
\begin{equation}
T^3\frac{\partial F}{\partial\xi}+\frac{\partial V(\phi
_a(m_\nu ))}{\partial m_\nu}=0 \ ,
\label{stationary2}
\end{equation}
where $\xi =m_\nu (T)/T$, $\rho _\nu =T^4F(\xi )$ and 
\begin{equation}
F(\xi )=\frac{1}{\pi ^2}\int _0^\infty \frac{dyy^2\sqrt{y^2+\xi
^2}}{e^y+1}\ .
\end{equation}
We have the time evolution of the neutrino mass from the relation of
 eq.(\ref{stationary2}). 
Since the stationary condition should be satisfied at the present epoch,
 the second constraint on the scalar potential is
\begin{equation}
\left[\left.T^3\frac{\partial F}{\partial\xi}+\frac{\partial V(\phi_a(m_\nu ))}{\partial m_\nu}\right]\right| _{m_\nu =m_\nu ^0,T=T_0}=0 \ .
\end{equation}
\noindent This condition turns to 
\begin{equation}
\left.\frac{\partial V(\phi_a(m_\nu ))}{\partial m_\nu}\right |_{m_\nu =m_\nu ^0,T=T_0}=-n^0_\nu \ ,
\end{equation}
where $n^0_\nu$ is the neutrino number density at the present epoch.

Since  neutrinos are supposed to be  non-relativistic at the present epoch, 
$\rho_\nu^0=m_\nu ^0n_\nu ^0$  is given and the equation of state becomes
\begin{equation}
w^0+1=\frac{m_\nu ^0n_\nu ^0}{m_\nu ^0n_\nu ^0+V(\phi _a^0(m_\nu^0))}\ .
\end{equation}
Taking the typical observed value $w_0=-0.9$,
 we can fix $\rho^0_\nu$. Then the neutrino mass $m^0_\nu$ is obtained
by putting the neutrino number density at the present epoch 
 $n_\nu ^0=8.82\times 10^{-13} \eV^3$ on $\rho^0_\nu= m^0_\nu n^0_\nu$.
Finally, we get  $m_\nu ^0=3.17 \ \eV$
 and 
$V(\phi_a^0(m_\nu^0))=2.52\times 10^{-11}\eV^4$,
where we take  $\rho_{\mbox{dark}}=0.7\rho_c=2.8\times 10^{-11}\eV^4$
at the present epoch.
 The neutrino mass $3.17 \ \eV$ may be large compared with the terrestrial  
neutrino experimental data. The neutrino mass of the $1\ \eV$ scale
is related with the LSND evidence \cite{LSND} and will be tested at the
MiniBOON experiment \cite{miniBOON}.
 On the other hand, putting $w_0=-0.998$ we get $m_\nu^0=0.05 \ \eV$,
which is  consistent with the atmospheric neutrino mass scale.
Thus, the value of $m_\nu^0$ depends on $w_0$.
In our following analyses, the numerical value of  $m_\nu^0$
is not so important as far as  the neutrino is non-relativistic 
at the present epoch. 
 We take  $m_\nu^0=3.17 \ \eV$ with  $w_0=-0.9$ as a reference value
in the following numerical studies. 

Now, we have two constraints on the potential and its derivative  
at the present epoch as follows:
\begin{eqnarray}
&&V(\phi _a^0(m_\nu ^0))=2.52\times 10^{-11}\hspace{2mm} \eV^4 \ ,
\label{data1}\\
\nonumber\\
&&\left.\frac{\partial V(\phi _a(m_\nu ))}{\partial m_\nu}\right|
_{m_\nu=m_\nu ^0}=-8.82\times 10^{-13}\hspace{2mm} \eV^3\ .
\label{data2}
\end{eqnarray}
It is found that the gradient of the scalar potential should be 
negative and very small. These constraints on the scalar potential are
very severe. 
 By using the potential of eq.(\ref{potential2}) in the model,
we have 
 \begin{eqnarray}
\frac{\partial V(m_\nu)}{\partial m_\nu}=\frac{1}{4\lambda^2} 
\left (m_\nu-\frac{m_D^2}{m_\nu}\right )^3  \left (1+\frac{m_D^2}{m_\nu^2}\right )+\frac{m_D^2}{2\lambda^2} 
\left (m_\nu-\frac{m_D^2}{m_\nu}\right )  \left (1+\frac{m_D^2}{m_\nu^2}\right ) \ .
 \end{eqnarray}
Therefore, the scalar potential satisfies the relation
 \begin{eqnarray}
  {\left.\frac{V(\phi _a(m_\nu))}{\frac{\partial V(\phi _a(m_\nu
  ))}{\partial m_\nu}}\right| _{m_\nu=m_\nu
 ^0~,~T=T_0}= \frac{m_\nu ^0}{4}
\frac{1-\frac{m_D^4}{(m_\nu^{0})^4}}{1+\frac{m_D^4}{(m_\nu^{0})^4}}> 
-\frac{m_\nu ^0}{4}} \ .
\label{PDP}
 \end{eqnarray}
This ratio  must  be  $-28.6$ from  eqs. (\ref{data1}) 
and (\ref{data2}), however,  our input $m_\nu^0=3.17 \ \eV$ 
  never  reproduce  this value. One cannot build any models with  only 
one superfield
$A$ unless the SUSY breaking term or the  $\mu$ term is added.

\subsection{$W=\frac{\lambda}{3}A^3+m_DLA$  with soft breaking terms}

Let us take into account the soft-breaking effect of the
supersymmetry. Then the scalar potential is given by
\begin{equation}
V(\phi _a)=\lambda ^2|\phi _a|^4+m_D^2|\phi _a|^2+m^2|\phi _a|^2+V_0 \ ,
\label{V0}
\end{equation}
where  $m$ is the soft-breaking mass and 
$V_0$ is a constant. 
The scale of the supersymmetry breaking in the standard sector 
$\tilde m$ is supposed to be  of order the electroweak scale $\tilde m=v$. 
However, if the dark sector couples to 
this supersymmetry breaking only via the neutrino,  radiative corrections
give the supersymmetry breaking mass scale of order $(m_D/v) \tilde m$.
Therefore, the soft-breaking mass $m$ in the dark sector 
is expected to be comparable to $m_D$, which is taken to be
 ${\mathcal O}(1 {\rm eV})$. 
Such a small soft mass of the supersymmetry breaking corresponds to 
 the small gravitino  mass  $m_{3/2}\simeq {\mathcal O}(1 {\rm eV})$,
which has been given  
in the gauge-mediated model of ref. \cite{Izawa}.

The gradient of the potential are given as 
\begin{equation}
\frac{\partial V(\phi _a(m_\nu ))}{\partial m_\nu}=
\frac{\phi_a}{\lambda} (2\lambda ^2\phi _a^2+m_D^2+m^2) \left
(1+\frac{m_D^2}{m_\nu^2}\right )\ ,
\end{equation}
\noindent where $\phi_a$ is given in terms of $m_\nu$  as in
eq.(\ref{mass}). Since we have four free parameters, $\lambda$, $m$,
$V_0$ and $m_D$, we can adjust  parameters to constraints of the
potential and its derivative. Putting the typical values for two
parameters by hand as follows:
\begin{equation}
\lambda =1 \ , \qquad m_D=10 \ eV \ ,
\end{equation}
with  $m_\nu^0=3.17 \ \eV$, we have $\phi^0_a(m_\nu^0 )=-14.2 \ \eV$. 
Then, $m^2$ and $V_0$ are fixed by the data of eqs.(\ref{data1}) and
(\ref{data2}) as follows:
\begin{eqnarray}
&&m^2=-2\times 14.2^2-10^2+\epsilon\hspace{2mm}(eV)^2\ ,
\label{fine1}\\
&&V_0=14.2^4-14.2^2\ \epsilon +0.7\rho _c-\rho _\nu
^0\hspace{2mm}(eV)^4 \ ,
\label{fine2}
\end{eqnarray}  
where $\epsilon =2\lambda ^2(\phi_a^{0})^2+m_D^2+m^2=-5.67\times
10^{-15}\ \eV^2$. It is remarked that the parameters  $m^2$ and $V_0$
are fine-tuned on the  order of $10^{-15} \eV^2$ to guarantee the tiny
$V(m_\nu)$ and $\partial V(m_\nu )/\partial m_\nu$ at the present
epoch, respectively. Using these parameters, we can get the evolution
of the neutrino mass and $w$ from the stationary condition and
the equation of state, respectively. However, since such a  case of
fine-tunings is not interesting, we do not discuss the case
furthermore.

\subsection{$W=\frac{\lambda}{3}A^3+m_DLA+\frac{\mu}{2}A^2$ model}

We consider the model including $\mu$-term as follows:
\begin{equation}
W=\frac{\lambda}{3}A^3+m_DLA+\frac{\mu}{2}A^2 \ ,
\end{equation}
which leads to the scalar potential as 
\begin{equation}
V(\phi _a)=|\lambda\phi _a^2+\mu\phi _a|^2+m_D^2|\phi _a|^2 \ .
\label{mupot}
\end{equation}
Taking a Lagrangian density of the form
\begin{equation}
{\mathcal L}=m_D\nu _L\psi _a+(2\lambda\phi _a+\mu )\psi _a\psi _a \ ,
\end{equation}
 we get $\phi_a$  in terms of the neutrino mass 
instead of eq.(\ref{mass}) as follows:
\begin{equation}
 \lambda \phi_a=\frac{m_\nu - \mu}{2}-\frac{m_D^2}{2 m_\nu}\ .
\label{massx}
\end{equation}
The derivative of the scalar potential is given as
\begin{eqnarray}
 \frac{\partial V(m_\nu)}{\partial m_\nu}= 
 \left (1+\frac{m_D^2}{m_\nu^2}\right )
 \frac{\phi_a}{\lambda}[(\lambda \phi_a+\mu)(2\lambda
 \phi_a+\mu)+m_D^2]\ .
 \end{eqnarray}
It is easily found that there is the parameter set, which satisfies
 the present data of eqs.(\ref{data1}) and (\ref{data2}) as
 \begin{equation}
  [(\lambda \phi_a+\mu)(2\lambda\phi_a+\mu)+m_D^2]\sim 10^{-10} \eV^2 \ ,
 \qquad m_D\ll\phi^0_a\sim \mu \sim 10^{-3} \eV\ \ .
\end{equation}
\noindent This result indicates the fine-tuning among $\lambda
 \phi_a$, $\mu$ and $m_D$ on the  order of $10^{-7}$. If the value of $m_D$ 
 is much smaller than values of $\lambda\phi _a$ and $\mu$, numerical
 solutions are
 \begin{equation}
  \phi_a=\pm 2.24\times 10^{-3}\eV \ , \quad \mu=\mp 4.48\times
  10^{-3}\eV \ ,
  \quad    |m_D|=10^{-4} \ \eV \ ,
 \end{equation}
\noindent with $\lambda=1$.
 Such a small value of $|\mu|\sim 10^{-3}\eV$ may be explained 
by the suppression of $M_{\rm TeV}^2/M_{\rm planck}$ \cite{Nomura}.

Since two mass eigenvalues are
almost degenerate due to  $m_D \gg 2\lambda\phi _a+\mu$,
 the left-handed neutrino $\nu_L$ mixes maximally with  $\psi_a$,
which is a kind of sterile neutrinos. Thus, this model is unfavored 
in the phenomenology of neutrino experiments.

\section{The double superfields model}
 It is very difficult to build a   model with the single superfield
 without fine-tuning of parameters of the  model.
 In this section, we introduce  two superfields, $A$ and $N$
 \cite{SI04}, which are singlets under the gauge group of the standard
 model.

\subsection{The simple model  $W=\lambda ANN+m_DLA+m_D'LN$}

It is assumed that the dark sector consists of two chiral superfields
$A$ and $N$, whose scalar and spinor components are $(\phi_a, \psi_a)$
and $(\phi_n, \psi_n)$, respectively, with the superpotential
\begin{equation}
W=\lambda ANN+m_DLA+m_D'LN \ ,
\end{equation}
\noindent
 where the scalar component $\phi_a$ of $A$ is assumed to be the
 acceleron. The scalar potential is given by
 \begin{equation}
  V(\phi _a,\phi _n)=\lambda ^2|\phi _n|^4+4\lambda ^2|\phi _a\phi
  _n|^2+m_D^2|\phi _a|^2+m_D'^2|\phi _n|^2
  \ .
  \label{potential3}
 \end{equation}
\noindent
The gradient of this potential is describing as follows:
 \begin{equation}
  \frac{\partial V(\phi _a)}{\partial\phi _a}=8\lambda ^2\phi _n^2\phi 
  _a+2m_D^2\phi_a \ ,
 \end{equation}
\noindent where  we assume the scalar component of two chiral
  superfields to be real. Then, we can write a Lagrangian density of
  the form
 \begin{equation}
  {\mathcal L}=m_D\nu _L\psi _a+m_D'\nu _L\psi _n+\lambda\phi _a\psi
  _n\psi _n+\lambda\phi _n\psi _a\psi _n\ .
 \end{equation}
Therefore the mass matrix in  the coupled system of the left-handed
  neutrino and the dark sector is given by
\begin{equation}
    \left(
    \begin{array}{ccc}
     0     & m_D            & m_D' \\
     m_D   & 0              & \lambda\phi _n \\
     m_D'  & \lambda\phi _n & \lambda\phi _a
    \end{array}
   \right) \  , 
\label{mass3}
 \end{equation}
\noindent
 in the $(\nu_L, \psi_a, \psi_n)$ basis.
The eigenvalue equation gives
 \begin{equation}
  \phi_a= \frac{m_\nu^3 -(\lambda ^2\phi
  _n^2+m_D^2+m_D'^2)m_\nu-2\lambda m_Dm_D'\phi _n}{m_\nu^2-m_D^2} \ .
 \end{equation}
\noindent By using this relation, the potential of
eq.(\ref{potential3}) is given in terms of the neutrino mass
 $m_\nu$. Putting $\lambda =1$ and $m_D'=1$ eV by hand, other three
 parameters are fixed by three constraints of $m_\nu ^0$, $V(m_\nu^0)$ and 
$\partial V(m_\nu )/\partial m_\nu |_{m_\nu =m_\nu ^0}$ as
 follows:
 \begin{equation}
  \phi _a^0=-2.42\times 10^{-15}\eV \ , \qquad |\phi _n|=5.02\times
  10^{-6}\eV \ , \qquad |m_D|=3.01\eV \ ,
 \end{equation}
where mass eigenvalues are obtained
 \begin{equation}
  m_\nu ^0=\pm 3.17\eV \ , \qquad -3.01\times 10^{-6}\eV \ .
 \end{equation}
We can see that two neutrinos are degenerate in the mass, in other
 words, $\nu _L$ and $\nu _a$ mix maximally. Therefore, this model is
 also unfabored in the phenomenology of the neutrino experiments.

\subsection{Right-handed neutrino}
Towards a realistic model, we introduce a right-handed heavy Majorana
neutrino, which is assumed to decouple from  $\psi_a$ and
$\psi_n$. Then, the effective mass matrix in eq.(\ref{mass3}) is
modified in the $(\nu_L, \psi_a, \psi_n)$ basis as follows:
 \begin{equation}
  \left(
   \begin{array}{ccc}
    C_{LL} & m_D            & m_D' \\
    m_D    & 0              & \lambda\phi _n \\
    m_D'   & \lambda\phi _n & \lambda\phi _a
   \end{array}
  \right) \  ,
  \label{mass4}
 \end{equation}
\noindent where $C_{LL}$ is the effective mass given by the seesaw
mechanism between the left-handed and right-handed neutrinos. The
eigenvalue equation gives
\begin{equation}
  \phi_a= \frac{m_\nu^3 -C_{LL}m_\nu^2-(\lambda^2\phi_n^2+m_D^2+m_D'^2)m_\nu
-2\lambda m_D m_D'\phi_n+C_{LL}\lambda^2 \phi_n^2}{m_\nu^2-m_D^2-C_{LL}m_\nu} \ .
\label{phia}
 \end{equation}
\begin{figure}
\begin{center}
\epsfxsize=8.  cm
\epsfbox{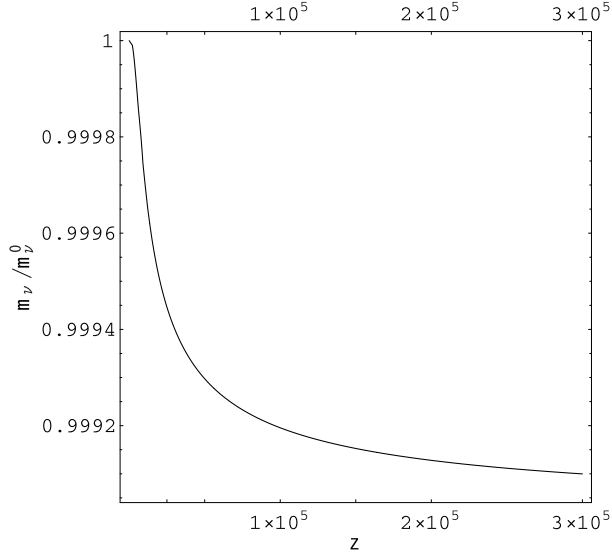}
\end{center}
\vspace*{-.1cm}
\caption{Plot of the scaled  neutrino mass versus the redshift $z$}
\end{figure}
Putting  $\lambda=1$, $m_D=0.01$eV and $m'_D=0.1\eV$ by hand, other
three parameters are fixed by the three constraints of $m^0_\nu$,
$V(m^0_\nu)$ and $\partial V(m_\nu )/\partial m_\nu|_{m_\nu=m_\nu^0}$
as follows:
\begin{equation}
\phi^0_a=-4.38\times 10^{-12} \eV \ , \quad \phi_n=\pm 5.02\times
10^{-5} \eV \ , \quad C_{LL}= 3.17 \ \eV \ , 
 \end{equation}
 where mass eigenvalues are obtained 
\begin{equation}
  m_\nu^0 =3.17\  \eV, \quad -3.17\times 10^{-3} \ \eV, \quad
 -9.18\times 10^{-6}  \ \eV \ ,
 \end{equation}
\noindent 
 where $3.17 \ \eV$ is the mass  of the active  neutrino,
and other ones are for sterile neutrinos.
Actually, the mixing between the active neutrino and sterile ones
are tiny.

The evolution of the neutrino mass is given  by using the stationary
condition of eq.(\ref{stationary2}). We show  the scaled  neutrino mass 
 $m_\nu/m^0_\nu$ versus the redshift $z=T/T_0-1$ in Fig.1,
because the absolute neutrino mass is not important as far as the neutrino
is non-relativistic at the present epoch.
As seen in Fig.1, the neutrino mass evolves  only $0.1\%$. 
This weak  $\phi_a$ dependence of the neutrino mass is understandable 
in the approximate mass formula:
\begin{eqnarray}
  m_\nu\simeq C_{LL}+\frac{m_D^2+m_D'^2+\lambda^2\phi_n^2}{C_{LL}}
  \left (1+\frac{1}{C_{LL}}\lambda\phi_a\right ) \ ,
\end{eqnarray}
\noindent where the constant term $C_{LL}=3.17\ \eV$ dominates the neutrino
mass and the $\phi_a$ dependence is suppressed on the  order of $m_D^2/C_{LL}^2$.

Once the evolution of $m_\nu$ is given, one can calculate
the  equation of state parameter $w$ as follows:
 \begin{equation}
  w+1=\frac{4- h(\xi)}{3\left[ 1+\frac{V(m_\nu)}{T^4 F(\xi)}\right ]} \ ,
 \end{equation}
\noindent  where
\begin{equation}
 h(\xi)=\frac{\xi}{F(\xi)}\frac{\partial F(\xi)}{\partial \xi} \ .
 \end{equation}
\begin{figure}
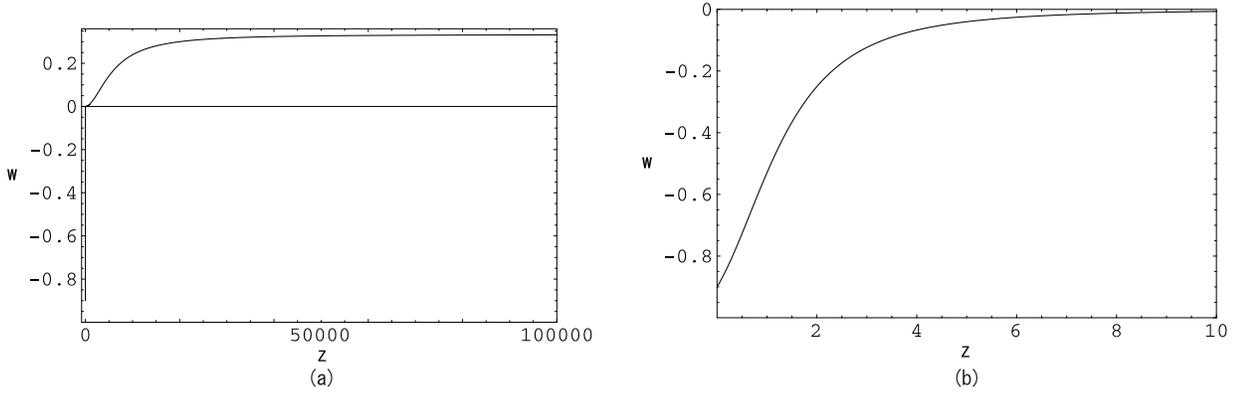

\begin{center}
\epsfxsize=7.8  cm
\epsfbox{dark2.ai}
\hskip 0.5 cm
\epsfxsize=7.8  cm
\epsfbox{dark3.ai}
\end{center}
\vspace*{-.1cm}
\caption{Plot of the equation of state parameter $w$ versus $z$
in the region of (a) $z=0\sim 100000$ and (b) $z=0\sim 10$.}
\end{figure}
The evolution of $w$ versus $z$ is shown in  Fig.2.
In order to see the behavior of $w$ near the present epoch,
we also plot $w$ at $z=0\sim 10$.
It is noticed that  $w$ evolves rapidly near the present epoch.

The evolution of the dark energy in the unit of $\rho_c$ is shown in Fig.3,
 in which the evolution
of the matter is presented in comparison. It is found that the matter 
dominates the energy density of the Universe  at $z\geq 1$.
\begin{figure}
\begin{center}
\epsfxsize=8.  cm
\epsfbox{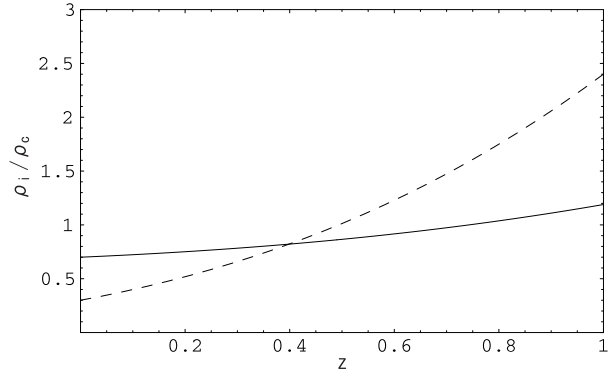}
\end{center}
\vspace*{-.1cm}
\caption{Plot of the energy density of the dark energy and of the matter 
in the unit of $\rho_c$ versus $z$, where the solid line and the dashed line
correspond to the dark energy and the matter, respectively.}
\end{figure}

In order to see when the acceleration of the cosmological expansion begun,
we calculate  the  acceleration  $\ddot a/a$ in the Friedmann equation;
\begin{eqnarray}
 \frac{\ddot a}{a}=-\frac{4\pi G}{3} [\rho_M + (3w+1)\rho_{\mbox{dark}}]\ ,
 \end{eqnarray}
\noindent where $\rho_M$ is the matter density and the contribution of
radiation is  neglected since we consider the epoch of $z=0\sim 1$.
As seen in Fig.4, the deceleration of the cosmological expansion
 converts to the acceleration  near $z=0.5$ in this model.
This result is different from the one in the power-law or exponential
potential discussed by Peccei \cite{Peccei}, in which
the conversion from the deceleration to the acceleration is predicted
 near $z=5\sim 7$.
\begin{figure}
\begin{center}
\epsfxsize=10.  cm
\epsfbox{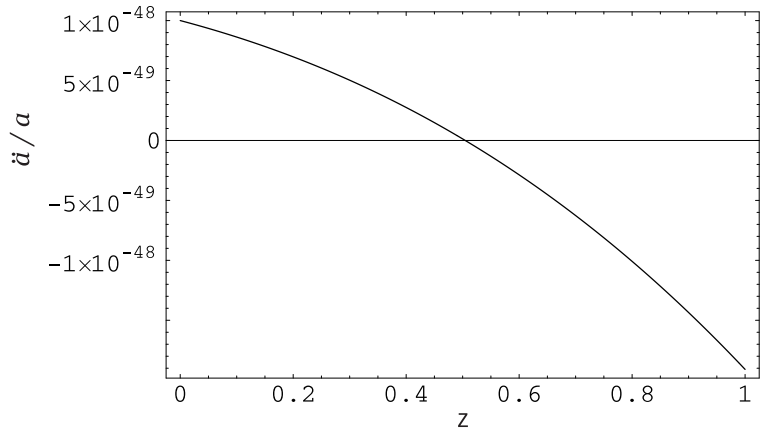}
\end{center}
\vspace*{-.1cm}
\caption{ Acceleration  $\ddot a/a$  versus $z$ }
\end{figure}
\section{Discussions and Summary }
In our work, we have presented numerical results 
in the case of the non-relativistic neutrino at the present epoch.
However, it was remarked that the speed of sound, $c_s$, which is given as
\cite{stability}
\begin{equation}
c_s^2=w+\frac{\dot w}{\dot\rho_{\mbox{dark}}}\ \rho_{\mbox{dark}}\ ,
 \end{equation}
\noindent   becomes imaginary in the non-relativistic limit at the present 
and then the Universe cease to accelerate.
Actually,  $c_s^2$ is negative in our potential  if we put $w_0=-0.9$,
which corresponds to $m_\nu^0=3.17\ \eV$.
On the other hand,   if we take $w_0=-0.998$, which leads to
$m_\nu^0=0.05 \ \eV$,  we get positive $c_s^2$  since
the time evolution of $\rho_{\rm dark}$  becomes slower 
in the case of  smaller  $m_\nu^0$ near the present epoch.
Therefore, the atmospheric mass scale of the neutrino mass $m_\nu^0=0.05\ \eV$
 may be  favored.

We have not discussed the quantum corrections to the scalar potential.
These corrections were discussed  in ref.\cite{Weiner}, in which  
it was remarked that the neutrino mass should be lower than $O(1\eV)$ 
although these are model dependent.
The quantum corrections will be investigated carefully 
in the coupled system of the neutrino and the  acceleron \cite{Doran}.

We have discussed the MaVaNs scenario in the supersymmetric theory and
found  a model which is  consistent with the cosmological data. 
 In the case of the model with the single superfield, one 
needs the soft SUSY breaking terms or the $\mu$ term. However,
 fine-tunings of some parameters are required to be consistent with 
 the cosmological data. 

In order to avoid these defects, we have discussed the model with 
two superfields, which is consistent with the  cosmological data. 
 However, the left-handed neutrino   mixes
 with the neutrino of the dark sector maximally in this case.

Adding a right-handed neutrino, which does not couple to the dark sector,
 we obtain the  model, in which  the mixing between  the left-handed neutrino and the neutrino of the dark sector is tiny. This model is
the first example of the MaVaNs with the supersymmetry.
In our  model, the deceleration of the cosmological expansion
 converts to the acceleration  near $z=0.5$.
The related phenomena of our scenario and the extension to the 
three families  of the active neutrinos 
will be discussed  elsewhere.

After this paper was submitted to the journal, 
the paper in ref. \cite{Weiner2} appeared,
 which presents a different supersymmetric model of MaVaNs.

 \section*{Acknowledgments}
We are most grateful to N. Weiner  for appropriate discussion at the
early stage of this work.
 We  thank the Yukawa Institute for Theoretical 
Physics at Kyoto University for support at the Workshop YITP-W-04-08 
(Summer Institute 2004, Fuji-Yoshida), where this research was initiated. 
  M. T is  supported by the Grant-in-Aid for Science Research,
 Ministry of Education, Science and Culture, Japan(No.16028205,17540243). 


\begin{thebibliography}{5}
\bibitem{Weiner}
R. Fardon, A. E. Nelson, N. Weiner, J. Cosmol. Astropart. Phys. {\bf 10},
 005 (2004).

\bibitem{Yanagida}
M. Kawasaki, H. Murayama and T. Yanagida,
 Mod. Phys. Lett. A {\bf 7}, 563 (1992).

\bibitem{McKellar}
G. J. Stephenson, T. Goldman and B. H. J. McKellar, 
Int. J. Mod. Phys. A {\bf 13}, 2765 (1998);  Mod. Phys. Lett. A {\bf 12}, 
2391 (1997).

\bibitem{Peccei}
R. D. Peccei, Phys. Rev. D {\bf 71}, 023527 (2005).

\bibitem{Kaplan}
D. B. Kaplan, A. E. Nelson, N. Weiner, Phy. Rev. Lett. {\bf 93}, 091801 (2004).

\bibitem{Mini}
V. Barger, D. Marfatia and K. Whisnant, hep-ph/0509163.

\bibitem{Wang}
P. Gu, X-L. Wang and X-Min. Zhang,  Phys. Rev. D {\bf 68}, 087301 (2003).

\bibitem{Bi}
X-J. Bi, P. Gu,  X-L. Wang  and X-Min. Zhang,   
  Phys. Rev. D {\bf 69}, 113007 (2004).

\bibitem{Gu}
 P. Gu and X-J. Bi,    Phys. Rev. D {\bf 70}, 063511 (2004).

\bibitem{Hung}
 P. Q. Hung and H. P\"as,  Mod. Phys. Lett. {\bf A20}, 1209 (2005).

\bibitem{SUN1}
V. Barger, P. Huber and D. Marfatia, hep-ph/0502196.

\bibitem{SUN2}
M. Cirelli and  M. C. Gonzalez-Garcia and C. Pe\~na-Garay, 
Nucl. Phys. {\bf B719}, 219 (2005).

\bibitem{Zhang}
X-J. Bi, B. Feng, H. Li and X-Min. Zhang,  hep-ph/0412002.

\bibitem{Right}
R. Horvat, astro-ph/0505507;\\
R. Barbieri, L. J. Hall, S. J. Oliver and A. Strumia,
Phys. Lett. {\bf B625} 189 (2005).

\bibitem{stability}
 N. Afshordi, M. Zaldarriaga and K. Kohri,  Phys. Rev. {\bf D72},
 065024 (2005).

\bibitem{WZ}
N. Weiner and K. Zurek,  hep-ph/0509201.

\bibitem{Supernova}
H. Li, B. Feng, J-Q. Xia and X-Min. Zhang,  hep-ph/0509272.

\bibitem{Data}
A. G. Riess $et.~al.$, Astrophys. J.{\bf 607}, 665 (2004).

\bibitem{LSND}
LSND Collaboration (A. Aguilan et al.), Phys. Rev. D {\bf 64}, 112007 (2001).

\bibitem{miniBOON}
MiniBooNE Collaboration (J. Monroe), hep-ex/0406048.

\bibitem{Izawa}
 K.I. Izawa, Prog. Theor. Phys. {\bf 98}, 443 (1997);\\
 K.I. Izawa and  T. Yanagida,
 Prog. Theor. Phys. {\bf 114}, 433 (2005).


\bibitem{Nomura}
Z. Chacko, L. J. Hall, Y. Nomura, JCAP 0410, 011 (2004).

\bibitem{SI04}
N. Weiner, Lecture at Summer Institute 2004 at Fuji-Yoshida (Japan),
 August 2004.

\bibitem{Doran}
For example, see 
M. Doran and J. J\"ackel, Phys. Rev. D {\bf 66}, 043519 (2002).

\bibitem{Weiner2}
R. Fardon, A. E. Nelson, N. Weiner,  hep-ph/0507235.

 
\end{thebibliography}
\end{document}